\documentclass[12pt,aps,prd,longbibliography,superscriptaddress,nofootinbib]{revtex4-2}
\pdfoutput=1
\usepackage[english]{babel}
\usepackage[utf8]{inputenc}
\usepackage[T1]{fontenc}
\usepackage{amsmath}
\usepackage{amssymb}
\usepackage{amsthm}
\usepackage{amsfonts}
\usepackage{url}
\usepackage{tensor}
\usepackage[bookmarks,colorlinks=false]{hyperref}
\usepackage{graphicx}
\usepackage{mathrsfs}
\usepackage{enumitem}
\usepackage{natbib}

\newcounter{mnotecount}[section]

\newcommand{\beq}{\begin{eqnarray}}
\newcommand{\eeq}{\end{eqnarray}}
\newcommand{\ben}{\begin{eqnarray*}}
\newcommand{\een}{\end{eqnarray*}}

\newtheorem*{theorem*}{Theorem}
\theoremstyle{definition}

\newcommand{\ha}{{\hat{\alpha}}}
\newcommand{\hb}{{\hat{\beta}}}
\newcommand{\hmu}{{\hat{\mu}}}
\newcommand{\hnu}{{\hat{\nu}}}
\newcommand{\hkappa}{{\hat{\kappa}}}
\newcommand{\hlambda}{{\hat{\lambda}}}

\begin{document}
\title{Freely falling bodies in a standing-wave spacetime}
\author{Sebastian J. Szybka}
\affiliation{Astronomical Observatory, Jagiellonian University}
\affiliation{Copernicus Center for Interdisciplinary Studies}
\author{Syed U. Naqvi}
\affiliation{Astronomical Observatory, Jagiellonian University}
%\date{}
\begin{abstract}
We study the motion of free masses subject to the influence of standing gravitational waves in the polarized Gowdy cosmology with a three-torus topology. We show that antinodes attract freely falling particles and we trace the velocity memory effect.
\end{abstract}
\maketitle{}

\section{Introduction}

Standing waves play a fundamental role in many branches of physics, but the nonlinearity of Einstein equations hinders their studies in gravitation (Bondi \cite{Hermann:2003}). The lack of a strict definition makes it difficult to go beyond an intuitive notion as to what gravitational standing waves are. Stephani tackled this problem in the paper \cite{Stephani:2003} where he presented an example of an exact standing-wave spacetime. An alternative definition was proposed in the paper \cite{Szybka:2019}. The intuitive notion of a standing wave implies that there is an alternating energy flow through nodes which averages to zero in time. Unfortunately, the energy of gravitational waves cannot be easily localized and the covariant averaging procedure is not known which makes a covariant definition of standing gravitational waves problematic. It has been suggested in the paper \cite{Szybka:2019} to evade this problem by taking the high frequency limit which captures the dominant contribution to the average energy flow. According to the definition presented in the paper \cite{Szybka:2019} this limit should correspond to an effective spacetime with the Ricci tensor of the Serge type $[(11)1,1]$. 

The aim of this paper is to clarify the behavior of test particles in a standing-wave spacetime. The standing-wave optical/acoustic traps, linear standing-wave particle accelerators, are commonly used in science and industry; thus, a relevant question is how gravitational standing waves influence trajectories of freely falling bodies.

The example of an exact standing-wave spacetime which was studied in the papers \cite{Stephani:2003,Szybka:2019} (see also \cite{Halilsoy:1988}), has a double interpretation. In the original form it belongs to the Einstein-Rosen class and possesses cylindrical symmetry. This solution and its high frequency limit have been also studied in details in the paper \cite{ers}. A simple transformation reveals that it may be reinterpreted as a particular case of the three-torus Gowdy model \cite{ers}.

A naive notion of concentration of the gravitational energy implies that the energy of standing gravitational waves is accumulated at antinodes because at antinodes the metric functions oscillate the most. This suggests that the geodesic equation should have stationary solutions at antinodes. One may show that this is not true for cylindrical standing waves (the Einstein-Rosen waves) discussed by Stephani \cite{Stephani:2003}. The exact solution studied there may be interpreted as a ``nonlinear'' superposition of incoming gravitational waves and their reflections from the symmetry axis. The amplitude of the waves decreases in radial direction. The highest concentration of the gravitational energy (as measured by Thorne's C-energy \cite{thorne})\footnote{See also the paper \cite{PhysRevD.55.669} where an alternative definition of the gravitational energy in cylindrical symmetry was given.} is at the symmetry center and the system effectively resembles a gravitational geon \cite{BH,Gyula:2017,Wheeler:1955}. However, this analogy is not strict. The gravitational energy is not confined to a finite region of space because of cylindrical symmetry. The test particles are kicked out from antinodes because the minimum of the gravitational ``potential'' is ``on average'' at the center. For a hypothetical external observer who studies the trajectories of freely falling bodies, the whole system looks like an infinite massive strut extended along the symmetry axis.\footnote{In contrast to the Brill-Hartle solution \cite{BH}, the high frequency limit of this solution is not globally regular \cite{ers}.} Therefore, this system is not similar to a standard setting in which electromagnetic or acoustic standing waves are investigated.

The Gowdy type interpretation of the standing-wave spacetime \cite{ers} is more promising here. The amplitude of gravitational waves does not change in spatial direction, but decreases in time as the universe expands.  The geodesic equation has stationary solutions at antinodes as it is shown in this paper. The motion of freely falling bodies may be studied relative to stationary observers at antinodes. We conduct such studies with the help of the geodesic deviation equation.

\section{Gowdy standing waves}\label{gsw}

The vacuum solution studied by Stephani \cite{Stephani:2003}  under a suitable transformation corresponds to a particular case of polarized Gowdy cosmologies with a three-torus topology \cite{ers}. The metric has the following form\footnote{In fact, $t$ should be substituted by $t/t_S$ in the metric \eqref{gowdy} and in the term $\ln{t}$ in $p$ [Eq.\ \eqref{ef}], where $t_S$ is a dimensional constant. Without loss of generality, we have assumed that $t_S=1$ in some not specified units (possibly {\it meters}). According to some terminology, the metric \eqref{gowdy} belongs to the generalized Einstein-Rosen class.}
\begin{equation}\label{gowdy}
	\hat g=e^f(-dt^2+dz^2)+t(e^pdx^2+e^{-p}dy^2)\;,
\end{equation}
where $0\leq z<2\pi$, $t>0$ ($t$ is a cosmic time function), $0\leq x,y<2\pi$, $f=f(t,z)$ and $p=p(t,z)$. The coordinates are ordered as $x^\alpha=(t,z,x,y)$. The particular solution we are interested in belongs to the class studied in the paper \cite{charachmalin} and is given by
\begin{equation}
\begin{split}\label{ef}
p=&-\ln{t}+2\beta\sqrt{\lambda}J_0(\frac{t}{\lambda})\sin(\frac{z}{\lambda})
	\;,\\
f=&\;\frac{\beta^2}{\lambda}t^2\left[J_0^2(\frac{t}{\lambda})+J_1^2(\frac{t}{\lambda})-2\frac{\lambda}{t}J_0(\frac{t}{\lambda})J_1(\frac{t}{\lambda})\sin^2(\frac{z}{\lambda})\right]\\
&-2\beta\sqrt\lambda J_0(\frac t\lambda)\sin(\frac{z}{\lambda})\;,
\end{split}
\end{equation}
where $J_i$ are the Bessel functions of the first kind and $i$th order. Periodicity of $f(t,0)=f(t,2\pi)$, $p(t,0)=p(t,2\pi)$ implies $\lambda=1/n$ with $n\in\mathbb{Z}$. The constant $\beta$ is arbitrary ($\beta=0$ corresponds to the Minkowski spacetime) and it controls the amplitude of the waves. Without loss of generality we assume $\beta>0$. There is a curvature singularity at $t=0$. (In polarized Gowdy models with all possible essential topologies $T^3$, $S^2\times S^1$, $S^3$, the strong cosmic censorship hypothesis holds \cite{Chrusciel_1990}.) After the big bang the model expands anisotropically. 

\section{Stationary observers at antinodes}\label{soa}

The geodesic equation \eqref{eq:geo} and the first integrals are presented in Appendix \ref{a1}. In order to study trajectories of test massive particles along the direction $\partial_z$, we investigate the geodesic equation under simplifying assumption $x=x_0$, $y=y_0$, where $x_0$, $y_0$ are constants. It takes a form
\begin{equation}
\begin{split}
	\nonumber
	\ddot t+ \frac{1}{2}\left(f_{,t} \dot{t}^2+2f_{,z} \dot t \dot z + f_{,t} \dot{z}^2\right)=&0\;,\\
	\ddot z+ \frac{1}{2}\left(f_{,z} \dot{z}^2+2f_{,t} \dot t \dot z + f_{,z} \dot{t}^2\right)=&0\;,
	\label{eg:geo}
\end{split}
\end{equation}
where overdots denote differentiation in a proper time $\tau$.
The first integral which follows from normalization of the four-velocity is
\begin{equation}\nonumber
	\dot{t}^2-\dot{z}^2=e^{-f}\;.
\end{equation}
We show below that there exists a stationary solution $\dot{z}=0$ ($z=z_0$, where $z_0$ is a constant which belongs to a particular set). The equations for $\dot{z}=0$ reduce to
\begin{eqnarray}
	0&=&\ddot{t}+\frac{1}{2}f_{,t}\dot{t}^2\;,\label{fc1}\\
	0&=&f_{,z}\;,\nonumber\\
	\dot{t}&=&e^{-\frac{1}{2}f}\;,\label{fc3}
\end{eqnarray}
where \eqref{fc3} is the first integral of \eqref{fc1}.
The single nontrivial condition is $f_{,z}=0$ which corresponds to antinodes. Using \eqref{ef}, we have $f_{,z}=(\dots)\cos{(z/\lambda)}$, thus $z=\lambda\pi(1/2+k)$, where $k\in\mathbb{Z}$, implies $f_{,z}=0$. Therefore, the curve $\gamma_k$, 
\begin{equation}\label{fge}
x^\mu=[t(\tau),\lambda\pi(1/2+ k),x_0,y_0]\;,
\end{equation} 
with $k\in\mathbb{Z}$ and $t(\tau)$ determined by \eqref{fc3} is a future-directed timelike geodesic and a stationary solution to the geodesic equation. Although odd and even $k$ lead to different values of the metric functions \eqref{ef} without loss of generality, we assume that $k=0$, unless explicitly indicated otherwise (the term $\sin(\frac{z}{\lambda})$ in these metric functions is multiplied by a function $J_0$ which is asymptotically periodic in $t$).

Behavior of massive particles around such stationary worldlines may be studied with the help of the geodesic deviation equation.

\section{Geodesic deviation equation}

 In order to avoid ``coordinate effects,'' it is convenient to study geodesic deviation equation in a nonholonomic basis. Let $(M,g)$ denote our spacetime with the metric given in the coordinate basis by \eqref{gowdy}. We introduce an orthonormal tetrad $\{e_{\hat \alpha}(p)\}\subset T_pM$
\begin{equation}\label{of}
\begin{split}
	e_{\hat 0}=&e^{-f/2}\partial_t\;,\\
	e_{\hat 1}=&e^{-f/2}\partial_z\;,\\
	e_{\hat 2}=&e^{-p/2}/\sqrt{t}\partial_x\;,\\
	e_{\hat 3}=&e^{p/2}/\sqrt{t}\partial_y\;.
\end{split}
\end{equation}
Its dual basis is denoted by $\{\theta^{\hat\alpha}(p)\}\subset T^*_p M$. The orthonormal tetrad $\{e_{\hat\alpha}\}$ is a freely falling frame of stationary observers at antinodes [along the geodesic $\gamma_k$ given by \eqref{fge}]. This fact is demonstrated with the help of the Cartan's first structure equation in Appendix \ref{a2}.

The geodesic deviation equation takes a particularly simple form in a freely falling frame. Let $\xi$ be a deviation vector, then along $\gamma_k$,
\begin{equation}\label{gde}
	\frac{d^2\xi^\ha}{d\tau^2}=-R^\ha_{\;\;\hat{0}\hb\hat{0}}\xi^\hb\;,
\end{equation}
where $\tau$ is a proper time of the observer. 

In Appendix \ref{a3}, we use the Cartan's second structure equation to find curvature two forms. Next, we determine the components of the Riemann tensor \eqref{Riemann}. It turns out that components $R^{\hat{\alpha}}_{\;\;\hat{0}\hat{\beta}\hat{0}}$ vanish for $\hat\alpha\neq\hat\beta$ so the equations decouple and only $R^{\hat{1}}_{\;\;\hat{0}\hat{1}\hat{0}}$, $R^{\hat{2}}_{\;\;\hat{0}\hat{2}\hat{0}}$, $R^{\hat{3}}_{\;\;\hat{0}\hat{3}\hat{0}}$ have to be taken into consideration. Substituting these components into the geodesic deviation equation \eqref{gde}, we obtain along $\gamma_k$ in terms of the coordinate $t$,\footnote{The equations have a simpler form in terms of coordinate time $t$ than in terms of the proper time of the observer $\tau$.}
\begin{equation}\label{fgd}
\begin{split}
\frac{d^2\xi^{\hat 1}}{dt^2}-\frac{1}{2}f_{,t}\frac{d\xi^{\hat 1}}{dt}&=\frac{1}{2}(f_{,tt}-f_{,zz})\xi^{\hat 1}=
-\frac{\beta}{\lambda t}J_1(\frac{t}{\lambda})\left[\sqrt\lambda+t\beta J_1(\frac{t}{\lambda})\right]\xi^{\hat 1}\;,\\
\frac{d^2\xi^{\hat 2}}{dt^2}-\frac{1}{2}f_{,t}\frac{d\xi^{\hat 2}}{dt}&=\frac{1}{4t}\left[f_{,t}+(p_{,t}(2-t f_{,t})+2t\; p_{,tt})\right]\xi^{\hat 2}\\
&=
-\frac{\beta}{\lambda^{3/2}}
\left[
J_0(\frac{t}{\lambda})-\frac{\lambda^2}{4\beta^2 t}
J_1^{-1}(\frac{t}{\lambda})f_{,t}^2
\right]
\xi^{\hat 2}\;,\\
\frac{d^2\xi^{\hat 3}}{dt^2}-\frac{1}{2}f_{,t}\frac{d\xi^{\hat 3}}{dt}&=\frac{1}{4t}\left[f_{,t}-(p_{,t}(2-t f_{,t})+2t\; p_{,tt})\right]\xi^{\hat 3}\\
&=
\frac{\beta}{\lambda^{3/2}}
\left[
J_0(\frac{t}{\lambda})-
\frac{\lambda}{2} J_1(\frac{t}{\lambda})f_{,t}
\right]
\xi^{\hat 3}\;,\\
\end{split}
\end{equation}
where $dt/d\tau=e^{-f/2}$ follows from the normalization of $e_{\hat 0}$ and where $$f_{,t}=2\frac{\beta}{\lambda} J_1(\frac{t}{\lambda})[\sqrt{\lambda}+t\beta J_1(\frac{t}{\lambda})]\;.$$ We have assumed without loss of generality $e_{\hat 0}\cdot \xi=0$; hence, $\xi^{\hat 0}=0$. In order to simplify the components $\hat 2$, $\hat 3$, we used one of Einstein equations $G_{\hat 0\hat 0}=0$ together with the fact that for our solutions \eqref{ef} we have $f_{,z}|_{\gamma_k}=0 \implies p_{,z}|_{\gamma_k}=0$. Equation \eqref{fgd} constitutes a decoupled system of autonomous second order linear ordinary differential equations with variable coefficients. Since the equation decouples, the motion of test particles in transverse and longitudinal directions can be studied separately.

The tidal forces in a proper time are directly related to the components of the Riemann tensor as may be seen in Eq.\ \eqref{gde}. In the freely falling frame $\{e_{\hat \alpha}\}$, the components $-R^{\hat{1}}_{\;\;\hat{0}\hat{1}\hat{0}}$, $-R^{\hat{2}}_{\;\;\hat{0}\hat{2}\hat{0}}$, $-R^{\hat{3}}_{\;\;\hat{0}\hat{3}\hat{0}}$ may be thought of as representing the gradient of the gravitational force in the instantaneous three-space of the stationary observers at antinodes. The positive values of components $R^{\hat{1}}_{\;\;\hat{0}\hat{1}\hat{0}}$, $R^{\hat{2}}_{\;\;\hat{0}\hat{2}\hat{0}}$, $R^{\hat{3}}_{\;\;\hat{0}\hat{3}\hat{0}}$ imply that test particles are attracted toward the observer in respective directions, and the negative values imply repulsion.

Equation \eqref{fgd} is quite complicated and we were unable to find exact solutions.
The geodesic deviation equation for $\xi^{\hat 1}$ at antinode may be written as
\begin{equation}\nonumber
\frac{d^2\xi^{\hat 1}}{dt^2}=-\frac{\beta}{\lambda t}J_1(\frac{t}{\lambda})\left[\sqrt\lambda+t\beta J_1(\frac{t}{\lambda})\right]\left(\xi^{\hat 1}-t\frac{d\xi^{\hat 1}}{dt}\right)\;.
\end{equation}
It may be integrated once; thus, one may present the solution in the integral form
\begin{equation}\nonumber
	\xi^{\hat 1}=\hat{a} t \int \frac{dt}{\lambda^2 t^2}e^{\frac{\beta^2}{2\lambda}t^2 \left[J_1(\frac{t}{\lambda})^2-J_0(\frac{t}{\lambda})J_2(\frac{t}{\lambda})\right]-\beta \sqrt{\lambda}J_0(\frac{t}{\lambda})}\;,
\end{equation}
where $\hat{a}$ is a constant. The second constant appears as a result of integration. Nevertheless, integral form of solution was not very helpful in our analysis.

Before turning to a numerical analysis, it is worth to obtain deeper understanding of the dynamics in the model. This may be achieved with the help of a null tetrad and the Newman-Penrose formalism \cite{Newman:1961qr}.

\subsection{Asymptotics and the Petrov type}

The standing gravitational waves studied in this paper may be thought of as a nontrivial superposition of two waves moving in the opposite spatial directions. In Appendix \ref{a4}, we introduced the Newman-Penrose tetrad with two null vectors $k^\mu$, $l^\mu$ aligned along direction of propagation of our waves. Next, we decomposed the Weyl scalar into the complex components $\psi_0,\dots,\psi_4$. The symmetries of the line element \eqref{gowdy} result in $\psi_1=\psi_3=0$. The physical interpretation of $\psi$'s \cite{gdevnh,CollidingPlaneWaves,exact2003} implies that $\psi_1$, $\psi_3$ represent longitudinal effects of waves in $k^\mu$ and $l^\mu$ directions, respectively. Similarly, $\psi_0$, $\psi_4$ represent the transverse effects of waves propagating in $k^\mu$, $l^\mu$ directions. The component $\psi_2$ is a Coulomb-like contribution to the geodesic deviation \cite{Szekeres1965}. Thus, in the context of the solution studied, the longitudinal forces acting on test particles in the geodesic deviation equation \eqref{gde} follow from the Coulomb part of the Weyl tensor $\psi_2$ not from the longitudinal effects of the waves $\psi_1=\psi_3=0$. The transverse forces are induced mainly by $\psi_0$ and $\psi_4$, but the Coulomb component $\psi_2$ also contributes to them.

The fact that $\psi_1=\psi_3=0$ allows us to apply technique outlined in the paper \cite{JanisNewman} to determine the Petrov type. Namely, we keep $l^\mu$ fixed and carry out null rotation which takes $k^\mu$ into a principal null vector. In the new tetrad $\psi'_0=0$. On the other hand, the transformation rules of $\psi_0$ are known, so
\begin{equation}\label{petrov}
\psi'=0=\psi_0+4 E\psi_1+6 E^2\psi_2+4 E^3\psi_3+E^4\psi_4\;,
\end{equation}
where $E$ is a complex parameter specifying the null rotation. The multiplicity of the roots of Eq.\ \eqref{petrov} encodes coincidences of principal null vectors which define the Petrov type. For $\psi_1=\psi_3=0$ the problem simplifies considerably: only four distinct or two double roots are possible which correspond to the Petrov types I and D, respectively. We define $\Delta=4(9\psi_2^2-\psi_0\psi_4)$. The analysis conducted with the help of the computer algebra system reveals that in general $\Delta\neq 0$ with $\Delta$ possibly vanishing on some hypersurfaces. Therefore, the studied spacetime is in general of the Petrov type I and it is not algebraically special.\footnote{Excluding the trivial flat solution for $\beta=0$ and the Kasner solution which belongs to the class represented by the line element \eqref{gowdy}, but requires generalization of the metric functions \eqref{ef}.}

The geodesic deviation equation \eqref{gde} could be written in terms of $\psi$'s. Using Eq.\ \eqref{psis} from Appendix \ref{a4}, we find $R^{\hat{1}}_{\;\;\hat{0}\hat{1}\hat{0}}=2\Psi_2$, $R^{\hat{2}}_{\;\;\hat{0}\hat{2}\hat{0}}=-\Psi_2+\frac{1}{2}\left(\Psi_0+\Psi_4\right)$, $R^{\hat{3}}_{\;\;\hat{0}\hat{3}\hat{0}}=-\Psi_2-\frac{1}{2}\left(\Psi_0+\Psi_4\right)$. At antinodes, $\Psi_4=\Psi_0$ and these equations reduce there to
\begin{equation}
\begin{split}\label{Rpsi}
R^{\hat{1}}_{\;\;\hat{0}\hat{1}\hat{0}}&=2\Psi_2\;,\\
R^{\hat{2}}_{\;\;\hat{0}\hat{2}\hat{0}}&=-\Psi_2+\Psi_0\;,\\
R^{\hat{3}}_{\;\;\hat{0}\hat{3}\hat{0}}&=-\Psi_2-\Psi_0\;.
\end{split}
\end{equation}
The Coulomb contribution $\Psi_2$ squashes a sphere of test particles into an ellipsoid deformed along $e_{\hat 1}$ direction. The transverse contribution $\Psi_0$ deforms a ring of test particles (in a plane stretched by vectors $e_{\hat 2}$, $e_{\hat 3}$) into an ellipse. 

The full formulas for $\Psi$'s are too long to be usefully presented here, but it is instructive to study their asymptotic behavior. 

In general, $\Psi$'s are almost periodic functions with alternating signs. For late times $t\gg\lambda$ we have 
\begin{equation}\nonumber
\Psi_0=O(1/\sqrt{t})\;,\qquad\Psi_4=O(1/\sqrt{t})\;,\qquad\Psi_2=O(1/t)\;.
\end{equation}
This asymptotics corresponds to the falloff condition of the Riemann tensor which is expected in the cosmological analogue \cite{Carmeli1985} of the cylindrical version of the famous peeling theorem \cite{Stachel1966}.\footnote{We note that Eqs.\ (10), (11), and (18) in the paper \cite{Carmeli1985} are inconsistent.} However, it should be noted that the standing waves in the cylindrical Einstein-Rosen model and in the Gowdy form do not satisfy the outgoing wave condition (nor an appropriate cosmological version of this condition). The standing waves contain the same amount of incoming and outgoing radiation, so different peeling properties of the Weyl tensor are expected. Precisely, for late times, $\Psi_0$ and $\Psi_4$ dominate the Weyl tensor and the solution looks like a superposition of two equal waves moving in opposite directions with the nonlinear artifact of such superposition, $\Psi_2$, becoming negligible.

Using formulas \eqref{psisA} from Appendix \ref{a4}, for late times $t\gg\lambda$ at antinodes $z=\lambda\pi(1/2+k)$, where $k\in\mathbb{Z}$,
\begin{equation}
\begin{split}\label{psisantinodes}
\Psi_{0}=\Psi_4\simeq& \sigma\frac{a e^{-a^2t}}{4}\frac{1}{\sqrt{t}}\left\{
\frac{2\sigma}{t}\cos\left(\frac{\pi}{4}+\frac{t}{\lambda}\right)-\frac{3a}{\sqrt t}\left[1-\sin\left(\frac{2t}{\lambda}\right)\right]
\right.\\&\left.+\frac{4}{\lambda}\sin\left(\frac{\pi}{4}+\frac{t}{\lambda}\right)+a^2\left[3\cos\left(\frac{\pi}{4}+\frac{t}{\lambda}\right)-\sin\left(\frac{\pi}{4}+3\frac{t}{\lambda}\right)\right]\right\}\;,\\
\Psi_2\simeq& \frac{a e^{-a^2t}}{4}\frac{1}{t}\left\{-\frac{2\sigma}{\sqrt t}\cos\left(\frac{\pi}{4}+\frac{t}{\lambda}\right)+a\left[1-\sin\left(\frac{2t}{\lambda}\right)\right]\right\}\;,
\end{split}
\end{equation}
where $a=\sqrt{\frac{2}{\pi}}\beta$ is a new auxiliary constant and $\sigma=1$ if $k$ is even, $\sigma=-1$ if $k$ is odd. 
The components $\Psi_0$, $\Psi_4$, $\Psi_2$ oscillate around $0$, but for $\Psi_2$ we have $$\Psi_2\gtrsim -\frac{a}{2\sqrt{2}\,t^{3/2}}e^{-a^2t}$$ and (for $m\in\mathbb{Z}$)
$$\max_{\pi\lambda(m-1)<t<\pi\lambda m}\Psi_2\gtrsim \frac{a^2}{2t}e^{-a^2t}\;.$$ 
The average value of $\Psi_2$ is above the zero. All this together with Eqs.\ \eqref{Rpsi} and \eqref{gde} imply that attraction to antinodes dominates along $e_{\hat 1}$. We will see in the next section that this fact remains true in quite early times. The freely falling particles are, on average, attracted to antinodes. This observation constitutes the main result of our work. We also note that the repelling contribution of $\Psi_2$ at antinodes along $e_{\hat 2}$, $e_{\hat 3}$ is overshadowed by dominant contribution of $\Psi_0=\Psi_4$. 

The number of waves on the three-torus is given by $n=1/\lambda$. We observe in Eqs.\ \eqref{psisantinodes} and \eqref{psisA} that for late times the magnitude of the Coulomb component $\Psi_2$ does not depend on a number of waves, but the magnitude of the transverse effect $\Psi_{0,4}$ does.

\begin{figure}[t!]
\begin{center}
%\label{fig:forces1}
\includegraphics[width=10cm,angle=0]{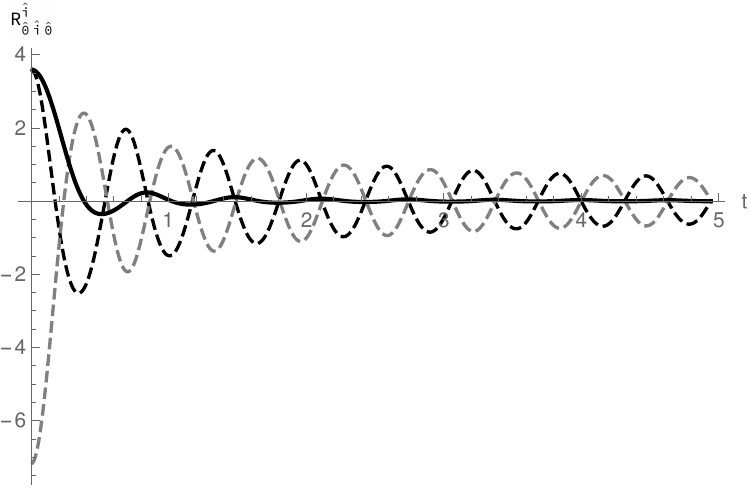}
\caption{The components of the Riemann tensor: $R^{\hat{1}}_{\;\;\hat{0}\hat{1}\hat{0}}$ (solid line), $R^{\hat{2}}_{\;\;\hat{0}\hat{2}\hat{0}}$ (dashed line), $R^{\hat{3}}_{\;\;\hat{0}\hat{3}\hat{0}}$ (dashed gray line) at the antinode which determine [via Eq.\ \eqref{gde}] the force acting on test particles along directions $\partial_z$, $\partial_x$, $\partial_y$, respectively. The positive values imply attraction, negative repulsion. (The parameters: $\lambda=1/10$, $\beta=1/5$.)}
\label{dirZ}
\end{center}
\end{figure}

For completeness, we provide asymptotics of $\psi$'s at antinodes for the early times $t\ll\lambda$,
\begin{equation}
\begin{split}\nonumber
\Psi_{0}=\Psi_4=&3\sigma \frac{e^{2\sigma\beta\sqrt{\lambda}}\beta}{4\lambda^{3/2}}+O(t^2)\;,\\
\Psi_2=&\sigma \frac{e^{2\sigma\beta\sqrt{\lambda}}\beta}{4\lambda^{3/2}}+O(t^2)\;.
\end{split}
\end{equation}
In terms of the components of the Riemann tensor we obtain $$2R^{\hat{1}}_{\;\;\hat{0}\hat{1}\hat{0}}+O(t^2)=2R^{\hat{2}}_{\;\;\hat{0}\hat{2}\hat{0}}+O(t^2)=-R^{\hat{3}}_{\;\;\hat{0}\hat{3}\hat{0}}+O(t^2)\;.$$
The behavior of the model is ``asymptotically velocity term dominated'' near the initial singularity.

\begin{figure}[t!]
\begin{center}
\includegraphics[width=10cm,angle=0]{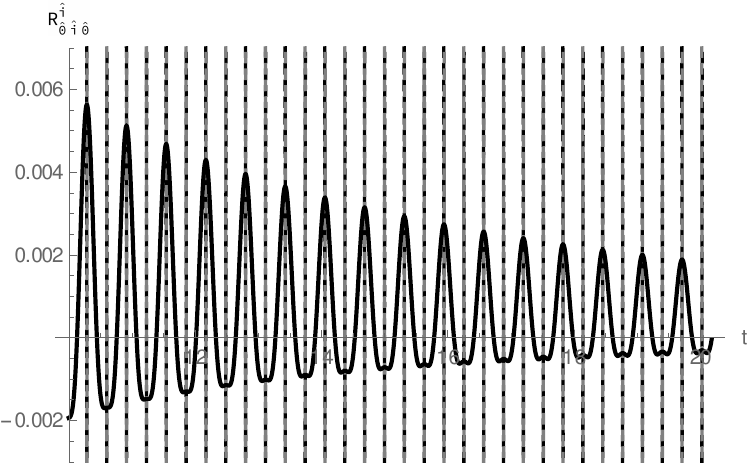}
\caption{The components of the Riemann tensor. A different range with all remaining details as in Fig.\ \ref{dirZ}. After initial phase, the transverse components of the force oscillate about zero with no clear preference for a sign (Fig.\ \ref{dirZ}). In contrast to that, the attraction toward the antinodes dominates in the longitudinal direction as the spacetime expands.}
\label{dirZ2}	
\end{center}
\end{figure}

Our analysis of the asymptotic behavior of the components of the Riemann tensor $R^{\hat{1}}_{\;\;\hat{0}\hat{1}\hat{0}}$, $R^{\hat{2}}_{\;\;\hat{0}\hat{2}\hat{0}}$, $R^{\hat{3}}_{\;\;\hat{0}\hat{3}\hat{0}}$ for late times and near the initial singularity may be verified with plots obtained with full formulas evaluated for a particular choice of parameters. Figures \ref{dirZ} and \ref{dirZ2} confirm our general considerations. 

%
%\clearpage

\subsection{Numerical analysis}

In this subsection, we present the numerical solutions to the geodesic deviation equation \eqref{fgd}. 

The linear gravitational waves are transverse; therefore, it is instructive to start our analysis with a ring of massive particles at one of the antinodes $z=\lambda(\frac{\pi}{2}+k)$, where $k\in \mathbb{Z}$. Without loss of generality, we choose $k=0$ which is equivalent in our formulas to all even values of $k$. The odd values of $k$ result in a slightly different equations, but for not too early times the behavior of metric functions is qualitatively similar. Analysis at odd antinodes may be thought of as a time-shifted analysis at even antinodes.

\begin{figure}[t!]
\begin{center}
\includegraphics[width=10cm,angle=0]{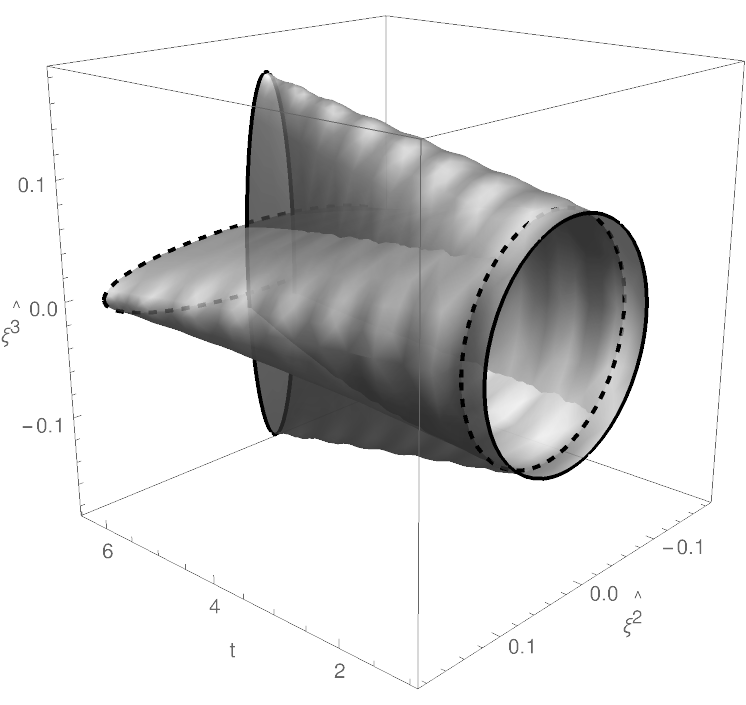}
\caption{The Tissot diagram. Two rings (solid and dashed) of test particles were initially at rest relative to the central observer in his freely falling frame. Since the solid ring corresponds to a different phase of the standing gravitational wave than of the dashed ring, then their evolution looks differently. The interaction with the first wave is decisive. The test particles will move along presented profiles through a caustic point (line) only slightly perturbed by subsequent waves. The long-range effect of the first wave on the bundle of test particles trajectories resembles the velocity memory effect. The parameters: $\lambda=1/10$, $\beta=1/5$, the radius at $t_0=1.173$ of the solid ring is $r=0.1$. The dashed ring of test particles is released at $t_0'=t_0+0.31$. The rings are evolved for $\Delta\tau=5$.}
\label{tissot}
\end{center}
\end{figure}

Since the model is expanding, then the amplitude of waves is decreasing in time. The initial conditions are crucial for the shape of the Tissot diagram---the interaction with the first wave determines the further evolution of the system. In Fig.\ \ref{tissot}, we see two rings (solid and dashed) of particles initially at rest  which evolve through a deformed ``corrugated tubes'' into ellipses (solid and dashed). These particles move on geodesics. They are initially (at some $t=t_0$) at rest relative to the stationary (central) observer at an antinode $d\xi^{\hat \alpha}/d\tau|_{t_0}=d\xi^{\hat \alpha}/dt|_{t_0}=0$. (This condition does not correspond to being stationary in our coordinate frame, so in general the test particles do not move along any $\gamma_k$.) The standing gravitational waves do not have a compact support, so it is not clear how to define the velocity memory effect in such spacetimes.\footnote{The memory effect for plane gravitational waves is described in the paper \cite{Zhang:2017rno}.} However, what we observe may be interpreted as some kind of the velocity memory: the trajectories of tests particles encode information about a phase of a standing gravitational wave at the initial time at which these particles have been released with zero initial velocity. This is clearly seen in Fig.\ \ref{tissot}---the long-range trajectories of particles in solid and dashed rings differ.

\begin{figure}[t!]
\begin{center}
\includegraphics[width=10cm,angle=0]{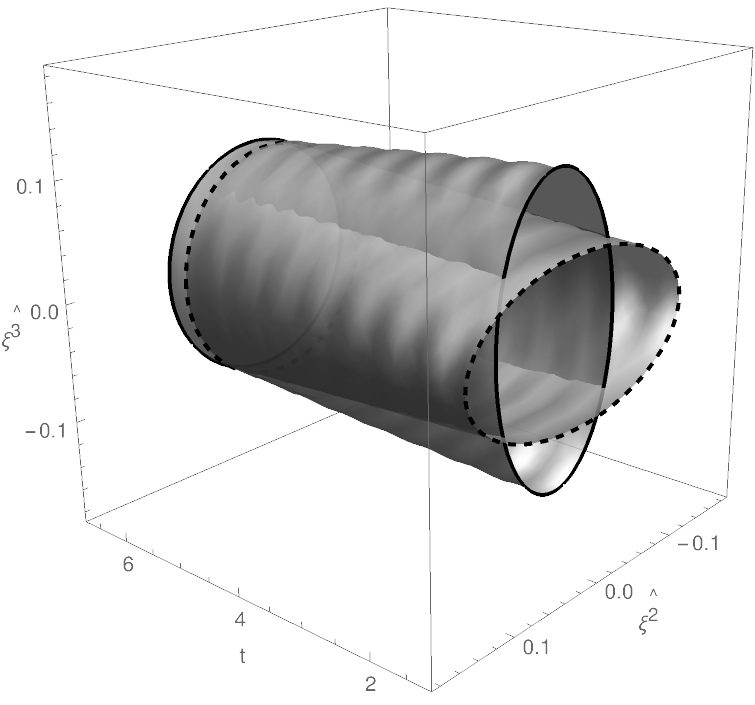}
\caption{The Tissot diagram for a contracting model ($t$ is a decreasing function of the proper time $\tau$). The parameters: $t_0=6.51$ (the solid ring), $t_0'=t_0-0.31$ (the dashed ring). Remaining details as in Fig.\ \ref{tissot}.}
\label{tissot2}
\end{center}
\end{figure}

The redefinition of the orientation of the time function does not alter our equations (only $dt/d\tau$ changes sign), but changes interpretation of the initial conditions: one may study the motion of test particles in the expanding universe with decreasing amplitude of standing waves or investigate the behavior of test particles in a contracting spacetime with emerging standing waves. The movement of test bodies in the contracting model is qualitatively similar to the behavior of free bodies in the expanding model (see Fig.\ \ref{tissot2}). The growing (in proper time of the observer at antinode) amplitude of gravitational waves leads to larger corrugation of ``tubes'' at the end of evolution (as compared to the expanding model). In both Figs.\ \ref{tissot} and \ref{tissot2}, the shift in the initial conditions alters a long-range behavior of trajectories (our analog of the velocity memory effect) and reveals the ``plus'' polarization of gravitational waves. 

\begin{figure}[t!]
\begin{center}
\includegraphics[width=10cm,angle=0]{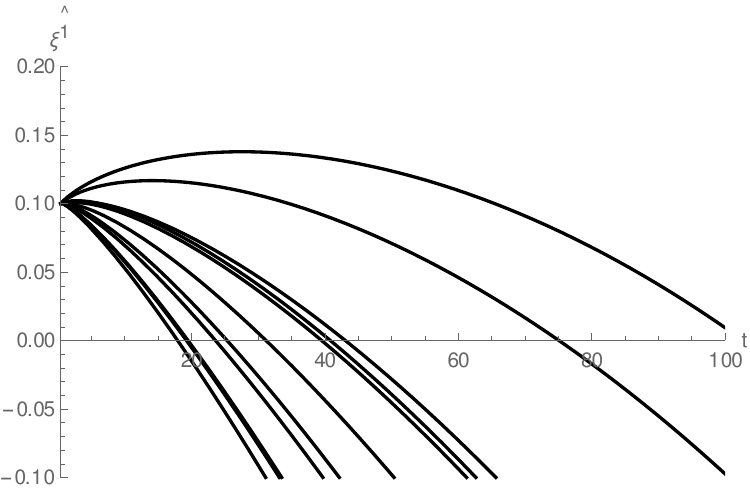}
\caption{
Trajectories of initially stationary test particles near antinode (the expanding model). Parameters: $\lambda=1/10$, $\beta=1/5$, $t_0$ varies uniformly from $0.38$ to $1.02$, $\xi^{\hat 1}(t_0)=0.1$, $d\xi^{\hat 1}/d\tau(t_0)=0$.
}
\label{dirZ1A}
\end{center}
\end{figure}

In order to study the motion of particles along $e_{\hat 1}$ direction (which corresponds in the coordinate frame to direction of $\partial_z$), we consider test particles with the single nonzero component of the deviation vector $\xi^{\hat 1}$. Similarly, as in case of the studies of the transverse effect described above, we consider test particles which are initially at rest relative to the stationary observer at an antinode $k=0$. The numerical analysis reveals that although some particles (it depends on the choice of $t_0$ and initial phase of the standing wave) may be initially repelled from the antinode, after some time the attraction dominates and particles move toward the antinode. This behavior is consistent with our analysis from the previous subsection and Figs.\ \ref{dirZ} and \ref{dirZ2}. Since the amplitude of the force decreases in the expanding model with $1/t$, then after initial interaction the particles move almost freely, pass the antinode, and fly away. This behavior, which may be interpreted as a memory effect of the first interaction with the wave, is presented in Fig.\ \ref{dirZ1A}, where trajectories of several particles starting at different $t_0$ were plotted.

\begin{figure}[t!]
\begin{center}
\includegraphics[width=10cm,angle=0]{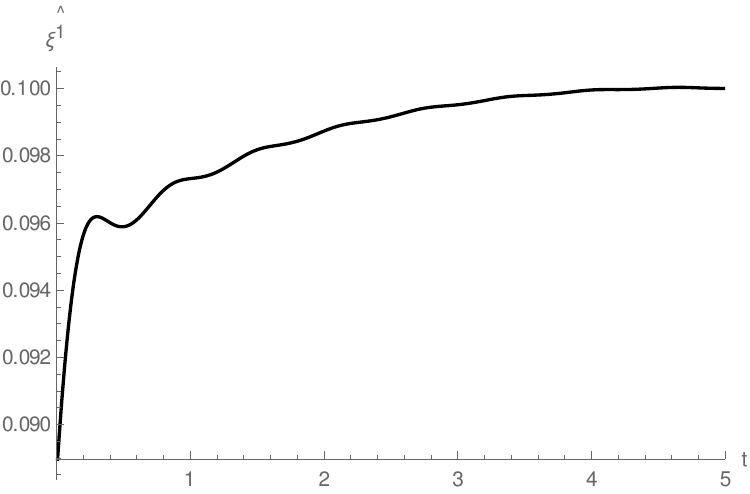}
\caption{
Trajectories of a typical initially stationary test particle near antinode in the contracting model. The initial time $t_0=5$. All remaining details as in Fig.\ \ref{dirZ1A}.
}
\label{dirZ2A}
\end{center}
\end{figure}

The behavior of freely falling bodies in the longitudinal direction changes if the collapsing model is examined. The typical trajectory of the initially stationary particle (at $t=t_0$) is presented in Fig.\ \ref{dirZ2A}. The choice of $t_0$ does not matter much this time. The particle is slowly attracted toward the antinode, but near the final crunch it starts to move more rapidly.

%\clearpage

\section{Geodesics}

The main result of our work has been presented in the previous sections where the geodesic deviation equation was investigated in the orthonormal freely falling frame of the stationary observer at an antinode (and also with the help of the Newman-Penrose null tetrad). In this section, in order to complete the research, we investigate the geodesic equation in a coordinate frame. Although strict interpretation of the results presented here is obscured by the coordinate dependence, they nicely fit to the global picture. We explore the structure of timelike geodesics to understand better behavior of freely falling bodies in the standing-wave spacetime. 

\begin{figure}[t!]
\begin{center}
\includegraphics[width=10cm,angle=0]{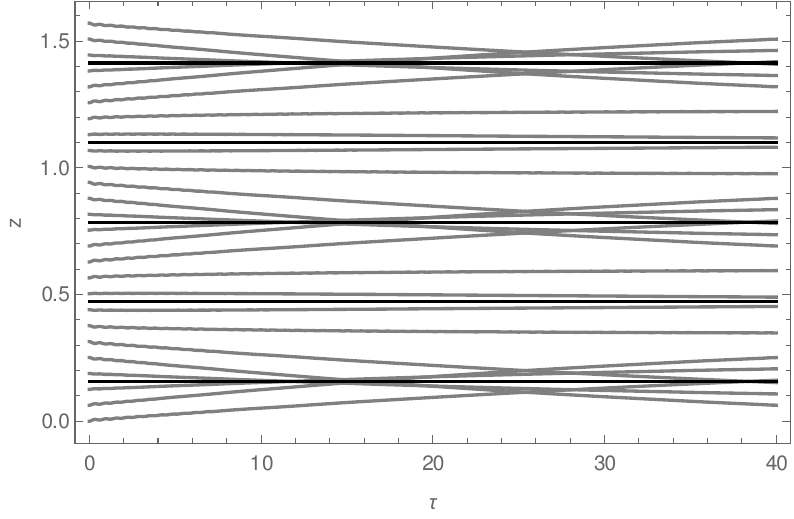}
\caption{Geodesics. Parameters: $\lambda=1/10$, $\beta=1/5$. The initial time $t(\tau=0)=0.1$. The final time $t(\tau=40)\simeq 32.4$.
}
\label{geodesics}
\end{center}
\end{figure}

\begin{figure}[t!]
\begin{center}
\includegraphics[width=10cm,angle=0]{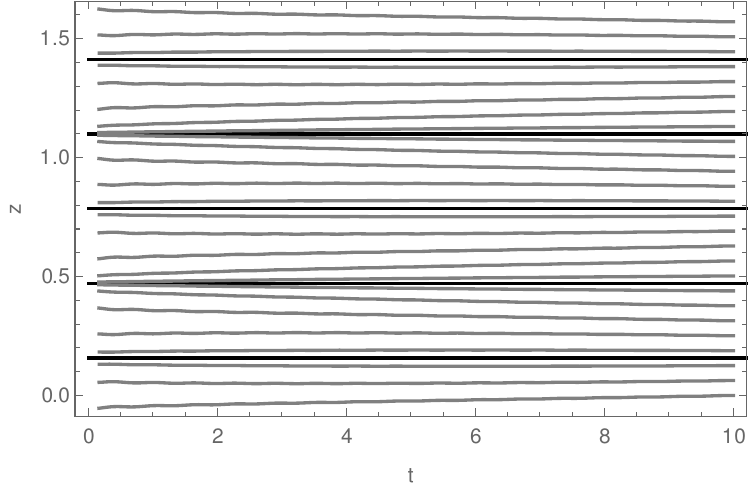}
\caption{Geodesics in contracting model. Parameters: $\lambda=1/10$, $\beta=1/5$. The initial time $t(\tau=0)=10$.
}
\label{geodesicsC}
\end{center}
\end{figure}

The geodesic equation is presented in Appendix \ref{a1}. In Sec.\ \ref{soa}, we proved that stationary observers at antinodes move along timelike geodesics given by the formula \eqref{fge}. In this section, we solve numerically the geodesic equation \eqref{eq:geo}. We consider $31$ test particles that were stationary in the coordinate frame at some initial moment $t_0$. The initial conditions are
\begin{equation}\label{icg}
\begin{split}
x^\alpha(\tau=0)&=(t_0,z_m,x_0,y_0)\;,\\
u&=e^{-f(t_0,z_i)/2}\partial_t\;,
\end{split}
\end{equation}
where $x_0$, $y_0$ are some constants that without loss of generality may be set to zero and where initial value of zero component $u^0$ of the particles' four-velocity $u$ follows from the normalization condition $u\cdot u=-1$. The values $z_m$ are uniformly spaced over several spatial periods of the metric functions, namely $z_m=2\pi/\lambda\;\frac{m}{10}$. The results are presented in Figs.\ \ref{geodesics} and \ref{geodesicsC}. The remaining parameters $\lambda$, $\beta$, and $t_0$ are given in the captions of these figures. Our choice of initial conditions \eqref{icg} together with the first integrals \eqref{fi} implies that the test particles do not move along $\partial_x$, $\partial_y$; thus, it remains to analyze their movement along the longitudinal direction $\partial_z$.

The geodesic equation is studied in the expanding (Fig.\ \ref{geodesics}) and contracting (Fig.\ \ref{geodesicsC}) model. The straight lines correspond to the test particles that were placed exactly at antinodes. The shape of trajectories is consistent with the analysis of the geodesic deviation equation. The antinodes attract test particles. This property is better visible in the expanding model because the amplitude of the gravitational waves is large at the initial time $t_0$. In the contracting model, the amplitude of the gravitational waves at the initial moment is small, so the attraction to antinodes is weaker. We point out that ``being at rest'' in the coordinate frame $\partial_\mu$ in general does not correspond to being at rest in our orthonormal frame $\{e_{\hat\mu}\}$; hence, our initial conditions \eqref{icg} for the geodesic equation are not the same as initial conditions for the deviation vector in the geodesic deviation equation. Nevertheless, the changes of the deviation vector in the expanding model (Fig.\ \ref{dirZ1A}) nicely correspond to what we observe for geodesics  (Fig.\ \ref{geodesics}). The similar analogy is valid in the contracting model (Figs.\ \ref{dirZ2A} and \ref{geodesicsC}).

%
%\clearpage

We see in Figs.\ \ref{geodesics} and \ref{geodesicsC} that there are two types of antinodes. They correspond to odd and even values of $k$ parameter which defines stationary geodesics $\gamma_k$ given by Eq.\ \eqref{fge}. The difference between these two types is not essential and their role may be reversed by a different choice of $t_0$.

In the expanding model, the presented geodesics cross the antinodes and pass hypothetical observers that were considered in our studies of the geodesic deviation. It is instructive to calculate the relative Lorentz factors and relative velocities.
 
The four-velocity of the stationary observer at the antinode in our coordinate basis is given by
\begin{equation}\nonumber
u_{obs}=e^{-f/2}\partial_t\;.
\end{equation}
The Lorentz factor of test particles as observed by stationary observers at antinodes is given by
\begin{equation}\nonumber
\gamma=u\cdot u_{obs}\;,
\end{equation}
where $u$ is a four-velocity of test particles. The numerical analysis reveals that for particles presented in Fig.\ \ref{geodesics} the velocities are nonrelativistic. They range from $0.0014 c$ to $0.0066 c$.

%\clearpage

\section{Summary}

%\vspace{0.3cm}

In this paper, we studied the behavior of freely falling bodies in a standing-wave spacetime. Using a particular example of such spacetime---the polarized three-torus Gowdy model---we showed that there exist stationary timelike geodesics at antinodes. We investigated the geodesic deviation equation at antinodes and proved that test particles are on average attracted to antinodes along the direction of the propagation of waves. This nicely confirms an intuitive notion of concentration of gravitational energy at places at which the geometry fluctuates at most. This phenomenon may be understood in terms of the components of the Weyl tensor in the Newman-Penrose tetrad which was adapted to the symmetries of the standing-wave spacetime.

Although our research is based on a particular exact solution, we think that the properties of standing waves discovered here are general. The Szekeres theorem \cite{Szekeres1965} implies that there do not exist vacuum solutions to Einstein equations with $\Psi_0\neq 0$, $\Psi_4\neq 0$, and $\Psi_1=\Psi_2=\Psi_3=0$. Two transverse gravitational waves cannot be trivially superposed. Therefore, the standing gravitational waves always induce additional effects which were represented in this work by the Coulomb component $\Psi_2$ and by attraction to antinodes. It is tempting to put forward a hypothesis that the Weyl tensor of any standing-wave spacetime contains non-zero, periodic, or almost periodic components $\Psi_0$, $\Psi_2$, $\Psi_4$ such that at antinodes in an appropriate Newman-Penrose tetrad $\Psi_0=\Psi_4$. Of course, our hypothesis needs further investigation.

\vspace{0.2cm}

\noindent{\sc Acknowledgments}

S.J.S.\ thanks Sasha Kovalska---the summer student---for a preliminary analysis of a problem which was related to our research. Some calculations were performed with the computer algebra system {\it Mathematica} and the {\sc xAct} package \cite{xAct}.

\clearpage

\appendix
\section{Geodesic equation}\label{a1}

The geodesic equation in coordinates $x^\alpha=(t,z,x,y)$ for the metric \eqref{gowdy} has a form
\begin{equation}
\begin{split}\label{eq:geo}
\ddot t &+\frac{1}{2}\left\{f_{,t} \dot{t}^2+2f_{,z}\dot t \dot z + f_{,t} \dot{z}^2+e^{-f}\left[e^{p}\dot{x}^2(1+p_{,t}t)+e^{-p}\dot{y}^2(1-p_{,t}t)\right]\right\}=0\;,\\
\ddot x &+\dot{x}(\dot{t}/t+p_{,t}\dot{t}+p_{,z}\dot{z})=0\;,\\
\ddot y &+\dot{y}(\dot{t}/t-p_{,t}\dot{t}-p_{,z}\dot{z})=0\;,\\
\ddot z &+\frac{1}{2}\left[f_{,z} \dot{z}^2+2f_{,t} \dot t \dot z + f_{,z} \dot{t}^2+t p_{,z}e^{-f}(-e^{p}\dot{x}^2+e^{-p}\dot{y}^2)\right]=0\;,
\end{split}
\end{equation}
where a dot denotes differentiation in the proper time $\tau$ or the affine parameter for timelike or null geodesics, respectively. The normalization of the four-velocity/wave vector gives rise to the first integral of the form
\begin{equation}\nonumber
-\epsilon=e^f(-\dot{t}^2+\dot{z}^2)+t(e^p\dot{x}^2+e^{-p}\dot{y}^2)\;,
\end{equation}
where the constant $\epsilon$ is equal to $1$ or $0$ for timelike or null geodesics, respectively. The Killing fields $\partial_x$, $\partial_y$ give two more quantities $c_x$, $c_y$ that are conserved along geodesics
\begin{equation}
\begin{split}\label{fi}
\dot x e^p t&=c_x\;,\\
\dot y e^{-p} t&=c_y\;.
\end{split}
\end{equation}

\section{Freely falling frame at antinodes}\label{a2}

We will show below that the orthonormal frame $\{e_{\hat\alpha}\}$ given by \eqref{of} corresponds to a freely falling frame of stationary observers at antinodes which move along the geodesics $\gamma_k$ with the tangent vector $u=e_{\hat 0}$ and $z=\lambda\pi(1/2+k)$, where $k\in\mathbb{Z}$. We have to show that along $\gamma_k$,
\begin{equation}\nonumber
\nabla_{e_{\hat 0}}e_{\hat\alpha}=0\; , \quad \nabla_{e_{\hat 0}}\theta^{\hat\alpha}=0\;.
\end{equation}
We have $\nabla_{e_{\hat 0}}e_{\hat\alpha}=\omega^{\hat\beta}_{\;\hat\alpha}(e_{\hat 0})e_{\hat\beta}$, where $\omega^{\hat\beta}_{\;\hat\alpha}$ are connection one forms. These one forms may be read out from the Cartan's first structure equation
\begin{equation}\label{cartan}
d\theta^{\hat\alpha}+\omega^{\hat\alpha}_{\;\hat\beta}\wedge \theta^{\hat\beta}=0\;.
\end{equation}
We have
\begin{equation}\nonumber
\begin{split}
d\theta^{\hat 0}=&-\frac{1}{2}e^\frac{f}{2}f_{,z}\;dt\wedge dz=-\frac{1}{2}e^{-\frac{f}{2}}f_{,z}\;\theta^{\hat 0}\wedge\theta^{\hat 1}\;,\\
d\theta^{\hat 1}=&\frac{1}{2}e^\frac{f}{2}f_{,t}\;dt\wedge dz=\frac{1}{2}e^{-\frac{f}{2}}f_{,t}\;\theta^{\hat 0}\wedge\theta^{\hat 1}\;,\\
d\theta^{\hat 2}=&\frac{1}{2}e^\frac{p}{2}\sqrt{t}\left[(\frac{1}{t}+p_{,t})\;dt\wedge dx+p_{,z}\;dz\wedge dx\right]\\
	=&\frac{1}{2}e^{-\frac{f}{2}}\left[(\frac 1 t +p_{,t})\;\theta^{\hat 0}\wedge\theta^{\hat 2}+p_{,z}\;\theta^{\hat 1}\wedge \theta^{\hat 2}\right] \;,\\
d\theta^{\hat 3}=&\frac{1}{2}e^{-\frac{p}{2}}\sqrt{t}\left[(\frac{1}{t}-p_{,t})\;dt\wedge dy-p_{,z}\;dz\wedge dy\right]\\
=& \frac{1}{2}e^{-\frac{f}{2}}\left[(\frac 1 t-p_{,t})\;\theta^{\hat 0}\wedge\theta^{\hat 3}-p_{,z}\;\theta^{\hat 1}\wedge \theta^{\hat 3}\right] \;.
\end{split}
\end{equation}
Comparing these equations with \eqref{cartan}, we find nonzero connection one forms
\begin{equation}\label{omegas}
	\begin{split}
		\omega^{\hat 0}_{\;\hat 1}&=\frac{1}{2}e^{-\frac{f}{2}}\left[f_{,z}\;\theta^{\hat 0}+f_{,t}\;\theta^{\hat 1}\right]\;,\\
		\omega^{\hat 0}_{\;\hat 2}&=\frac{1}{2}e^{-\frac{f}{2}}(1/t+p_{,t})\;\theta^{\hat 2}\;,\\
		\omega^{\hat 0}_{\;\hat 3}&=\frac{1}{2}e^{-\frac{f}{2}}(1/t-p_{,t})\;\theta^{\hat 3}\;,\\
		\omega^{\hat 1}_{\;\hat 2}&=-\frac{1}{2}e^{-\frac{f}{2}}p_{,z}\;\theta^{\hat 2}\;,\\
		\omega^{\hat 1}_{\;\hat 3}&=\frac{1}{2}e^{-\frac{f}{2}}p_{,z}\;\theta^{\hat 3}\;.
	\end{split}
\end{equation}
Since $f_{,z}|_{\gamma_k}=0$, then $\nabla_{e_{\hat 0}}e_{\hat\alpha}=0$. This (together with the cobasis duality relation and the Lebniz rule for the covariant derivative) implies $\nabla_{e_{\hat 0}}\theta^{\ha}=0$. To sum up, our nonholonomic basis $\{e_{\ha}\}$ is a freely falling frame of the observers moving along geodesics $\gamma_k$ (observers at antinodes). 

\section{Derivation of the geodesic deviation equation}\label{a3}

In the freely falling frame, not all components of the Riemann tensor are needed to derive the geodesic deviation equation. We consider the freely falling frame $\{e_{\hat\alpha}\}$ given by \eqref{of} and its dual basis denoted as $\{\theta^{\hat\alpha}\}$.

The Cartan's second structure equation gives the curvature two forms $\Omega^{\hat\alpha}_{\;\hat\beta}$,
\begin{equation}\label{cartan2}
\Omega^{\hat\alpha}_{\;\hat\beta}=d\omega^{\hat\alpha}_{\;\hat\beta}+\omega^{\hat\alpha}_{\;\hat\sigma}\wedge\omega^{\hat\sigma}_{\;\hat\beta}\;.
\end{equation}
Since $\Omega^{\hat\alpha}_{\;\hat\beta}=\frac{1}{2}R^{\hat\alpha}_{\;\hat\beta\hat\sigma\hat\delta}\theta^{\hat\sigma}\wedge\theta^{\hat\delta}$, then not all curvature two forms are needed.
Using the connection one forms \eqref{omegas}, we find
\begin{equation}\nonumber
\begin{split}
\Omega^{\hat 0}_{\;\hat 1}&=\frac{1}{2}e^{-f}(f_{,tt}-f_{,zz})\theta^{\hat 0}\wedge\theta^{\hat 1}\;,\\
\Omega^{\hat 0}_{\;\hat 2}&=\frac{1}{4}e^{-f}\left[(p_{,t}+\frac{1}{t})^2-f_{,t}(p_{,t}+\frac{1}{t})+2(p_{,tt}-\frac{1}{t^2})-f_{,z}p_{,z}\right]\theta^{\hat 0}\wedge\theta^{\hat 2}\\&+\frac{1}{4}e^{-f}\left[(p_{,t}+\frac{1}{t})(p_{,z}-f_{,z})-f_{,t}p_{,z}+2p_{,tz}\right]\theta^{\hat 1}\wedge\theta^{\hat 2}\;,\\
\Omega^{\hat 0}_{\;\hat 3}&=\frac{1}{4}e^{-f}\left[(p_{,t}-\frac{1}{t})^2+f_{,t}(p_{,t}-\frac{1}{t})-2(p_{,tt}+\frac{1}{t^2})+f_{,z}p_{,z}\right]\theta^{\hat 0}\wedge\theta^{\hat 3}\\&+\frac{1}{4}e^{-f}\left[(p_{,t}-\frac{1}{t})(p_{,z}+f_{,z})+f_{,t}p_{,z}-2p_{,tz}\right]\theta^{\hat 1}\wedge\theta^{\hat 3}\;.
\end{split}
\end{equation}
One may read out the necessary components of the Riemann tensor
\begin{equation}
\begin{split}\label{Riemann}
R^{\hat 1}_{\;\hat 0\hat 1\hat 0}&=\frac{1}{2}e^{-f}(f_{,zz}-f_{,tt})\;,\\
R^{\hat 2}_{\;\hat 0\hat 2\hat 0}&=-\frac{1}{4}e^{-f}\left[(p_{,t}+\frac{1}{t})^2-f_{,t}(p_{,t}+\frac{1}{t})+2(p_{,tt}-\frac{1}{t^2})-f_{,z}p_{,z}\right]\;,\\
R^{\hat 3}_{\;\hat 0\hat 3\hat 0}&=-\frac{1}{4}e^{-f}\left[(p_{,t}-\frac{1}{t})^2+f_{,t}(p_{,t}-\frac{1}{t})-2(p_{,tt}+\frac{1}{t^2})+f_{,z}p_{,z}\right]\;.
\end{split}
\end{equation}

The metric functions take form ($\sigma=1$ for $k\in 2\mathbb{Z}$ and $\sigma=-1$ for $k\in 2\mathbb{Z}+1$)
\begin{equation}\nonumber
\begin{split}
p|_{\gamma_k}&=-\ln{t}+2\sigma\beta\sqrt{\lambda}J_0(\frac{t}{\lambda})\;,\\
f|_{\gamma_k}&=\frac{\beta^2}{\lambda}t^2\left[J_0^2(\frac{t}{\lambda})+J_1^2(\frac{t}{\lambda})-2\frac{\lambda}{t}J_0(\frac{t}{\lambda})J_1(\frac{t}{\lambda})\right]-2\sigma\beta\sqrt\lambda J_0(\frac t\lambda)\;,\\
f_{,t}|_{\gamma_k}&=\frac{2\beta}{\lambda}J_1(\frac{t}{\lambda})\left[\sigma\sqrt\lambda+t\beta J_1(\frac{t}{\lambda})\right]\;.
\end{split}
\end{equation}

\section{Components of the Weyl tensor in the Newman-Penrose tetrad}\label{a4}

It is convenient for the analysis of gravitational waves to decompose the Weyl tensor into the Newman-Penrose tetrad components. Following Stephani {\it et al.}\ \cite{exact2003}, we adapt the original notation of Newman and Penrose \cite{Newman:1961qr} to the signature $+2$. We introduce a complex null tetrad $w_{\breve{\mu}}=\{k,l,m,\bar m\}$ using the orthonormal tetrad $\{e_\ha\}$ as follows,
\begin{equation}
\begin{split}\label{tetrad}
k&=\frac{1}{\sqrt 2}(e_{\hat 0}+e_{\hat 1})\;\quad l=\frac{1}{\sqrt 2}(e_{\hat 0}-e_{\hat 1})\;,\\
m&=\frac{1}{\sqrt 2}(e_{\hat 2}-i e_{\hat 3})\;\quad \bar m=\frac{1}{\sqrt 2}(e_{\hat 2}+i e_{\hat 3})\;,
\end{split}
\end{equation}
where two real vectors $k$, $l$ and two complex conjugate vectors $m$, $\bar m$ are null. These vectors' inner products vanish except
\begin{equation}\nonumber
k\cdot l=-1\;,\quad m\cdot\bar m=1\;.
\end{equation}
The components of the metric with respect to this null complex tetrad are $g_{\breve{\alpha}\breve{\beta}}=-2k_{(\breve{\alpha}}l_{\breve{\beta})}+2m_{(\breve{\alpha}} \bar m_{\breve{\beta})}$. The Weyl tensor has ten independent components which are determined by the five complex coefficients $\Psi_0$, $\Psi_1$, $\Psi_2$, $\Psi_3$, $\Psi_4$. 

The choice of a null tetrad \eqref{tetrad} is not unique and may be changed by Lorentz transformations of the frame. Since we want to adapt the Newman-Penrose null tetrad to the problem at hand, then we assume an additional condition. The standing wave may be seen as a nontrivial superposition of gravitational waves moving in opposite spatial directions along $e_{\hat 1}$ (which is one of the spatial vectors of our orthonormal freely falling frame). Therefore, we assume that $e_{\hat 1}\cdot k=-e_{\hat 1}\cdot l$. This condition does not define the tetrad uniquely, but is sufficient for the purpose of this study.

For the vacuum spacetimes $C_{\hkappa\hlambda\hmu\hnu}=R_{\hkappa\hlambda\hmu\hnu}$. We find for the spacetime studied in this paper, 
\begin{equation}
\begin{split}\nonumber
\Psi_0&:=C_{\hkappa\hlambda\hmu\hnu}k^\hkappa m^\hlambda k^\hmu m^\hnu=C_{\breve{0}\breve{2}\breve{0}\breve{2}}\\
&=\frac{1}{4}\left(C_{\hat{0}\hat{2}\hat{0}\hat{2}}-C_{\hat{0}\hat{3}\hat{0}\hat{3}}+C_{\hat{1}\hat{2}\hat{1}\hat{2}}-C_{\hat{1}\hat{3}\hat{1}\hat{3}}\right)+\frac{1}{2}\left(C_{\hat{0}\hat{2}\hat{1}\hat{2}}-C_{\hat{0}\hat{3}\hat{1}\hat{3}}\right)\\
\Psi_1&:=C_{\hkappa\hlambda\hmu\hnu}k^\hkappa l^\hlambda k^\hmu m^\hnu=C_{\breve{0}\breve{1}\breve{0}\breve{2}}
=0\;,\\
\Psi_2&:=C_{\hkappa\hlambda\hmu\hnu}k^\hkappa m^\hlambda {\bar m}^\hmu l^\hnu=C_{\breve{0}\breve{2}\breve{3}\breve{1}}=\frac{1}{4}\left(-C_{\hat{0}\hat{2}\hat{0}\hat{2}}-C_{\hat{0}\hat{3}\hat{0}\hat{3}}+C_{\hat{1}\hat{2}\hat{1}\hat{2}}+C_{\hat{1}\hat{3}\hat{1}\hat{3}}\right)\\
\Psi_3&:=C_{\hkappa\hlambda\hmu\hnu}l^\hkappa k^\hlambda l^\hmu {\bar m}^\hnu=C_{\breve{1}\breve{0}\breve{1}\breve{3}}
=0\;,\\\
\Psi_4&:=C_{\hkappa\hlambda\hmu\hnu}l^\hkappa {\bar m}^\hlambda l^\hmu {\bar m}^\hnu=C_{\breve{1}\breve{3}\breve{1}\breve{3}}\\
&=\frac{1}{4}\left(C_{\hat{0}\hat{2}\hat{0}\hat{2}}-C_{\hat{0}\hat{3}\hat{0}\hat{3}}+C_{\hat{1}\hat{2}\hat{1}\hat{2}}-C_{\hat{1}\hat{3}\hat{1}\hat{3}}\right)-\frac{1}{2}\left(C_{\hat{0}\hat{2}\hat{1}\hat{2}}-C_{\hat{0}\hat{3}\hat{1}\hat{3}}\right)\;.
\end{split}
\end{equation}
We have $C_{\hat{0}\hat{3}\hat{1}\hat{3}}=-C_{\hat{0}\hat{2}\hat{1}\hat{2}}$, $C_{\hat{1}\hat{3}\hat{1}\hat{3}}=-C_{\hat{0}\hat{2}\hat{0}\hat{2}}$, $C_{\hat{1}\hat{2}\hat{1}\hat{2}}=-C_{\hat{0}\hat{3}\hat{0}\hat{3}}$, and the Weyl tensor is traceless; hence,
\begin{equation}
\begin{split}\label{psis}
\Psi_0&=\frac{1}{2}(C_{\hat{0}\hat{2}\hat{0}\hat{2}}-C_{\hat{0}\hat{3}\hat{0}\hat{3}})+C_{\hat{0}\hat{2}\hat{1}\hat{2}}=\frac{1}{2}(R^{\hat{2}}_{\;\hat{0}\hat{2}\hat{0}}-R^{\hat{3}}_{\;\hat{0}\hat{3}\hat{0}})+C_{\hat{0}\hat{2}\hat{1}\hat{2}}\;,\\
\Psi_4&=\frac{1}{2}(C_{\hat{0}\hat{2}\hat{0}\hat{2}}-C_{\hat{0}\hat{3}\hat{0}\hat{3}})-C_{\hat{0}\hat{2}\hat{1}\hat{2}}=\frac{1}{2}(R^{\hat{2}}_{\;\hat{0}\hat{2}\hat{0}}-R^{\hat{3}}_{\;\hat{0}\hat{3}\hat{0}})-C_{\hat{0}\hat{2}\hat{1}\hat{2}}\;,\\
\Psi_2&=-\frac{1}{2}\left(C_{\hat{0}\hat{2}\hat{0}\hat{2}}+C_{\hat{0}\hat{3}\hat{0}\hat{3}}\right)=\frac{1}{2}C_{\hat{0}\hat{1}\hat{0}\hat{1}} =\frac{1}{2}R^{\hat{1}}_{\;\hat{0}\hat{1}\hat{0}}\;,
\end{split}
\end{equation}
where components of the Riemann tensor are given by Eq.\ \eqref{Riemann} and where $C_{\hat{0}\hat{2}\hat{1}\hat{2}}=O(1/\sqrt{t})$ in general, but $C_{\hat{0}\hat{2}\hat{1}\hat{2}}=0$ at antinodes. Therefore, at antinodes, $\Psi_0=\Psi_4$ holds.
For late times $t\gg \lambda$, we obtain
\begin{equation}
\begin{split}\label{psisA}
\Psi_{0,4}\simeq&\mp \frac{a e^{-a^2t}}{4}\frac{1}{\sqrt{t}}\left\{\mp \frac{2}{t} \cos\left(\frac{\pi}{4}+\frac{t}{\lambda}\right)\sin\left(\frac{z}{\lambda}\right)\pm\frac{3a}{\sqrt{t}}\left[1+\sin\left(2\frac{t\pm z}{\lambda}\right)\right]\right.\\&\left.+ \frac{4}{\lambda}\cos\left(\frac{\pi}{4}+\frac{t\pm z}{\lambda}\right)+a^2\left[\cos\left(\frac{\pi}{4}+3\frac{t\pm z}{\lambda}\right)-3\sin\left(\frac{\pi}{4}+\frac{t\pm z}{\lambda}\right)\right]\right\}\;,\\
\Psi_2\simeq& -\frac{a e^{-a^2t}}{4}\frac{1}{t}\left\{\frac{2}{\sqrt{t}}\cos\left(\frac{\pi}{4}+\frac{t}{\lambda}\right)\sin\left(\frac{z}{\lambda}\right)+a\left[\cos\left(\frac{2z}{\lambda}\right)+\sin\left(\frac{2t}{\lambda}\right)\right]\right\}\;,
\end{split}
\end{equation}
where $a=\sqrt{\frac{2}{\pi}}\beta$ and the top and bottom signs correspond to $\Psi_0$, $\Psi_4$, respectively. 

\bibliography{report.bib}

\end{document}